\documentclass[10pt,a4paper,twoside]{article}
\usepackage{epsfig}
\usepackage{baltlat6}
\usepackage{array}
\usepackage{amssymb}
\begin{document}
\ \
\vspace{0.5mm}

\newcommand{\me}{\mathop{\sf Me}\nolimits}
\setcounter{page}{387}

\titlehead{Baltic Astronomy, vol.\,24, 387--394, 2015}

\titleb{ON THE POSSIBILITY OF DETERMINING THE DISTANCE TO THE GALACTIC
CENTER FROM THE GEOMETRY OF SPIRAL ARM SEGMENTS}

\begin{authorl}
\authorb{I. I. Nikiforov}{} and
\authorb{A. V. Veselova}{}
\end{authorl}

\moveright-3.2mm
\vbox{
\begin{addressl}
\addressb{}{Department of Celestial Mechanics, Saint~Petersburg State University,\\
Universitetskij Prospekt 28,
Staryj Peterhof, Saint~Petersburg 198504, Russia; nii@astro.spbu.ru}
\end{addressl}  }

\submitb{Received: 2015 November 2; accepted: 2015 November 30}

\begin{summary}
A new approach to determining the solar galactocentric distance,
$R_0$, from the geometry of spiral-arm segments is proposed.
Geometric aspects of the problem are analyzed and a simplified
three-point method for estimating $R_0$ from objects in a spiral
segment is developed in order to test the proposed approach. An
estimate of $R_0 = 8.44 \pm 0.45$~{kpc} is obtained by applying the
method to masers with measured trigonometric parallaxes, and
statistical properties of the $R_0$ estimation from spiral segments
are analyzed.
\end{summary}

\begin{keywords} Galaxy: structure -- Galaxy: fundamental
parameters -- masers
\end{keywords}

\resthead{Distance to the Galactic center from spiral arm segments} {I.
I. Nikiforov, A. V. Veselova}

\sectionb{1}{INTRODUCTION}

The study of the spiral structure of the Milky Way is of fundamental
importance for galactic astronomy and dynamics. Till recently,
spatial modeling of tangent and other concentrations of gas and
young objects -- one of techniques used in the field -- was
principally based on the {\em positions\/} of spiral-arm segments
(short as a rule) rather than on the {\em geometry\/} of individual
segments. Unfortunately, this approach yields discrepant results
(cf., e.g., Efremov 2011; Francis \& Anderson 2012). However, recent
high-precision measurements of heliocentric distances (parallaxes)
for fair tracers of {\em long\/} segments of spiral arms provide the
possibility of spatial modeling of individual spiral segments (e.g.,
Xu et al.\/ 2013; Reid et al.\/ 2014; Bobylev \& Bajkova 2014).

The latter approach is a direct method for estimating the pitch
angles for the segments (see, e.g., Bobylev \& Bajkova 2014).
However, we can take a step further trying to treat the distance to
the center of the Galaxy, $R_0$, as a free parameter in spatial
modeling of spiral segments. In case of success, we gain (1)~a more
comprehensive modeling given the close interrelation between the
pitch angle and $R_0$ (according to our calculations) and (2)~a new
method for determining  $R_0$ from objects of the flat Galactic
subsystem. The proposed technique is by nature a spatial and
absolute (if it uses maser parallaxes) method of $R_0$ measurement
(see Nikiforov 2004).

In this paper, we consider geometric aspects of the problem
(Section~2), apply our new approach to masers with trigonometric
parallaxes (Section~3), and examine numerically the statistical
properties of the $R_0$ estimation from spiral segments (Section~4).
\enlargethispage{3mm}

\sectionb{2}{DETERMINATION OF THE GEOMETRIC PARAMETERS OF THE\\
LOGARITHMIC SPIRAL FROM POINTS OF ITS SEGMENT}

\vskip1mm

We assume that the spiral arm is a logarithmic spiral with a
constant pitch angle~$i$. The Galactic center is assumed to be
situated at the pole of the spiral and the solar galactocentric
distance $R_0$ is considered to be equal to the~distance to this
pole. The galactoaxial distance~$R$ of a~point on the~logarithmic
spiral is defined by the~equation
\begin{equation}
 R(\lambda) = |R_0|e^{k(\lambda - \lambda_0)},
\end{equation}
where $\lambda \in  ( - \infty; + \infty )$  is the \textit{rotary}
galactocentric longitude of the point and $k \equiv \tan{i}$,
$\lambda_0$ is the positional parameter. Longitude $\lambda$ is
counted clockwise as seen from the North Galactic Pole; $\lambda = 0
\pm 2\pi{n}$, $n \in \mathbb{Z}$, in the direction to the Sun. We
suppose that the~direction to the Galactic center is known. Although
trailing spiral arms are expected, we accept solutions irrespective
of whether the resulting pitch angle is negative or positive.

First we consider the problem on the number of points that define
the logarithmic spiral as a geometric figure. Given that the
logarithmic spiral has three parameters ($R_0$, $i$, and
$\lambda_0$) we try to~determine them from three points $M_i$,
$i=1,2,3$, supposed to belong to the spiral. We solve the problem in
terms of the {\em nominal\/} galactocentric longitudes, $\Lambda_i$,
which specify the positions of points on the Galactic plane. The
longitudes $\Lambda_i \in ( -\pi; +\pi],  i = 1, 2, 3$, can~be found
from the following equations:
  \begin{equation}\label{X_i_La_i}
 \sin{\Lambda_i} = Y_i/R_i,\quad \cos{\Lambda_i} = (R_0 - X_i)/R_i\,,
    \end{equation}
    \begin{equation}
      R_i^2 = R_0^2 +r_i^2\cos^2 b_i  - 2R_0r_i\cos b_i\cos l_i\,,
    \end{equation}
where $r_i$ is the heliocentric distance of the $i$th point; $l_i$
and $b_i$ are its Galactic longitude and latitude respectively;
$X_i$ and $Y_i$ are the Cartesian heliocentric coordinates, which
are defined by the equations
\begin{equation}
        X_i = r_i\cos b_i\cos l_i\,, \quad  Y_i = r_i\cos b_i\sin l_i\,.
\end{equation}
The $X$-axis points toward the Galactic center and the $Y$-axis
points in the direction of Galactic rotation. The Sun is located at
the origin of the coordinate system.

The solution of the problem is given by the following equations:
 \begin{equation}
 (\Lambda_3 - \Lambda_2)\ln{R_1} + (\Lambda_1 - \Lambda_3)\ln{R_2} + (\Lambda_2 - \Lambda_1)\ln{R_3} = 0; \label{basic_ln}
  \end{equation}
\begin{equation}
   k = \ln\left(R_i/R_j\right)\big/({\Lambda_i - \Lambda_j}),\quad i, j = 1,2,3, \enskip i \ne j; \label{k_}
\end{equation}
\begin{equation}
  \lambda_0 = \Lambda_i - \ln\left(R_i/|R_0|\right)/k, \quad i = 1, 2, 3
  \label{l_0}.
\end{equation}
The parameter $R_0$ can be determined from Equation (5), and $k$ and
$\lambda_0$ can then be calculated from Equations (6) and (7),
respectively. Given the transcendental nature of Equation~(5), below
we place an emphasis on the number of its roots.

Here we use the model spiral with $R_0 = 8~$kpc, $i =-18\fdg7$,
$\lambda_0 = -30\fdg0$ (Nikiforov \& Shekhovtsova 2001) to
illustrate our findings. In the case where all points $M_i$,
$i=1,2,3$, are located on the same side of the $X$-axis,
Equation~(5) has always two roots and hence two spirals pass through
the three points (see the left panel in Fig.~1). In the case where
the points are on different sides of the $X$-axis, Equation~(5) can
have one, two or three roots, however,  in most of the cases there
is only one unique root (see Fig.~1, right panel). The equation has
three roots if $(\Lambda_3 - \Lambda_1) \gtrsim 100^\circ$ and
$|\Lambda_2| \lessapprox (\Lambda_3-\Lambda_1)/2$ (see Fig.~2). Note
that, since there is always one root equal to the initial value
($R_{0,1}=8$~kpc), the total number of roots for given
$\Delta\Lambda$ and $\Lambda_2$ is determined by the number of
intersections of the line $\Lambda_2=\mathrm{const}$ with the branch
of additional roots for this $\Delta\Lambda$ in Fig.~2. One or two
such intersections mean two or three roots, correspondingly; if
there is no intersection, the root $R_{0,1}$ is the only one.


\begin{figure}[!tH]
\vbox{
\hspace*{2pt}%
\centerline{\psfig{figure=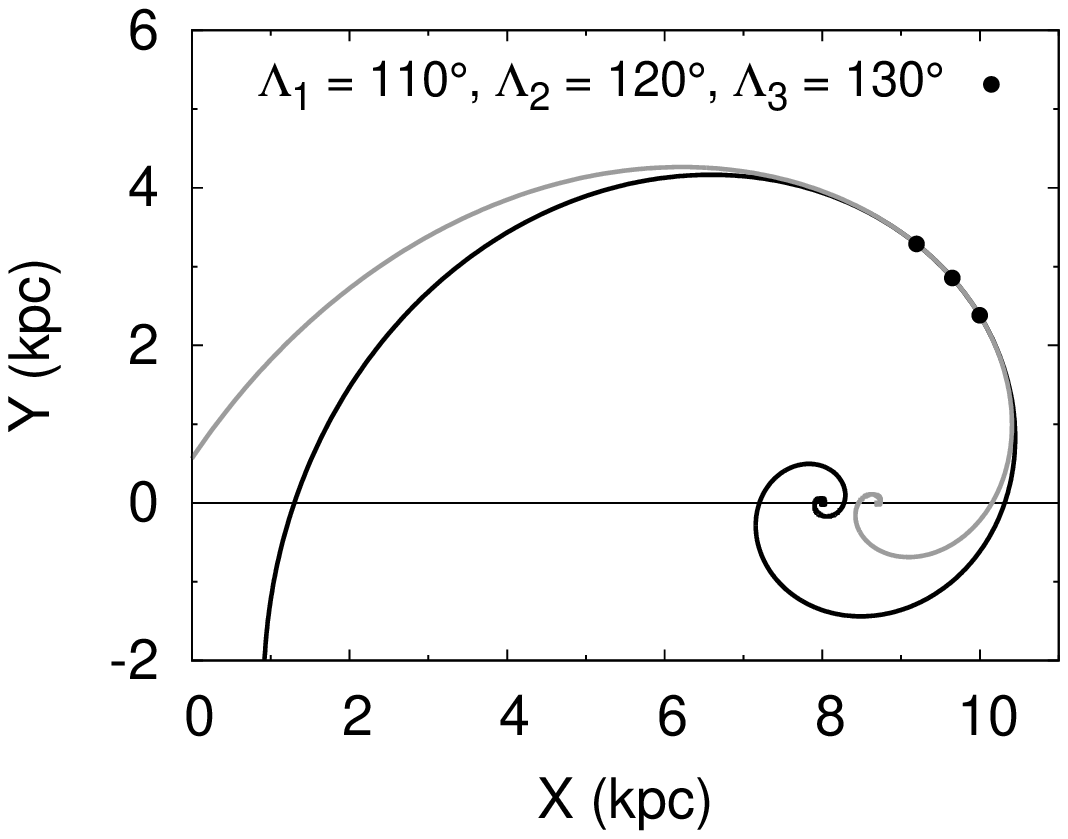,width=64mm,angle=0,clip=}%
\raisebox{0.0pt}{%
\hspace*{-15pt}%
\psfig{figure=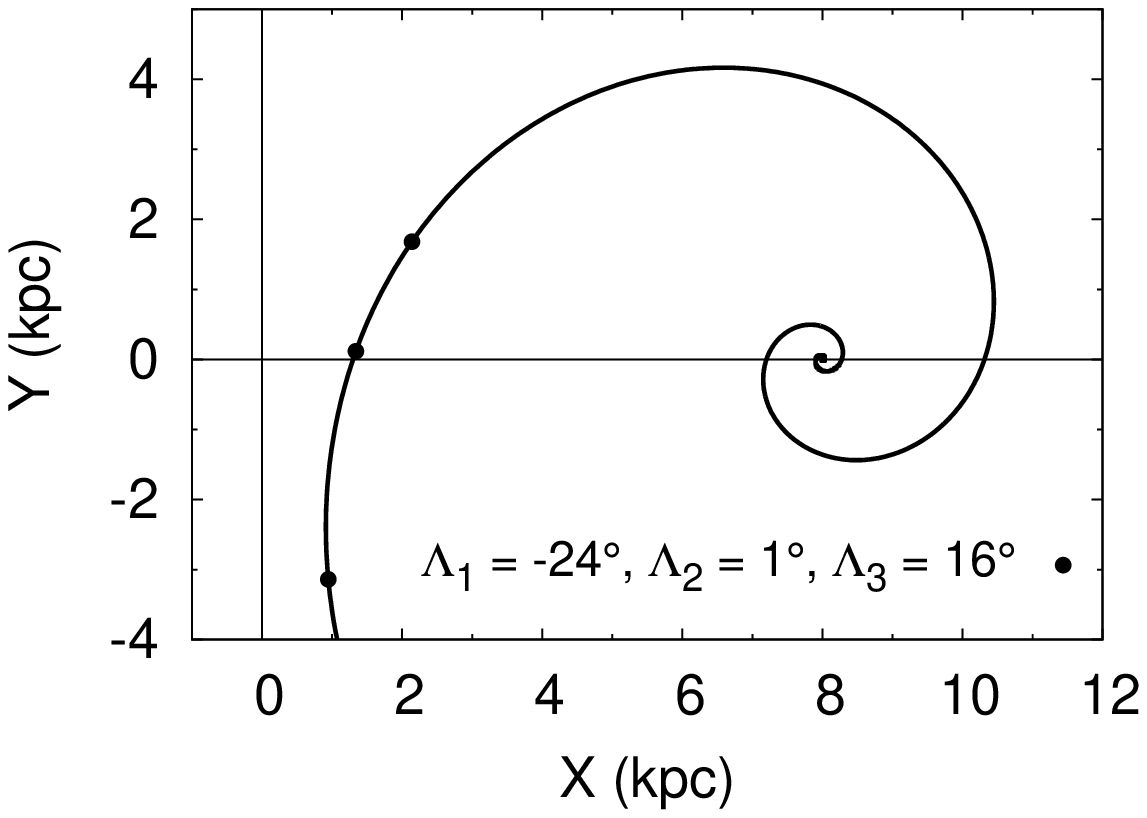,width=64mm,angle=0,clip=}%
}}
\captionb{1}{Spirals passing through the three given points. The
black line shows the model spiral.  The gray line (left panel) shows
the second spiral which passes through the given points in the case
if they are located on the same side of the $X$-axis.} }\vspace{5pt}
\end{figure}


\begin{figure}[!tH]
\vbox{ \centerline{\psfig{figure=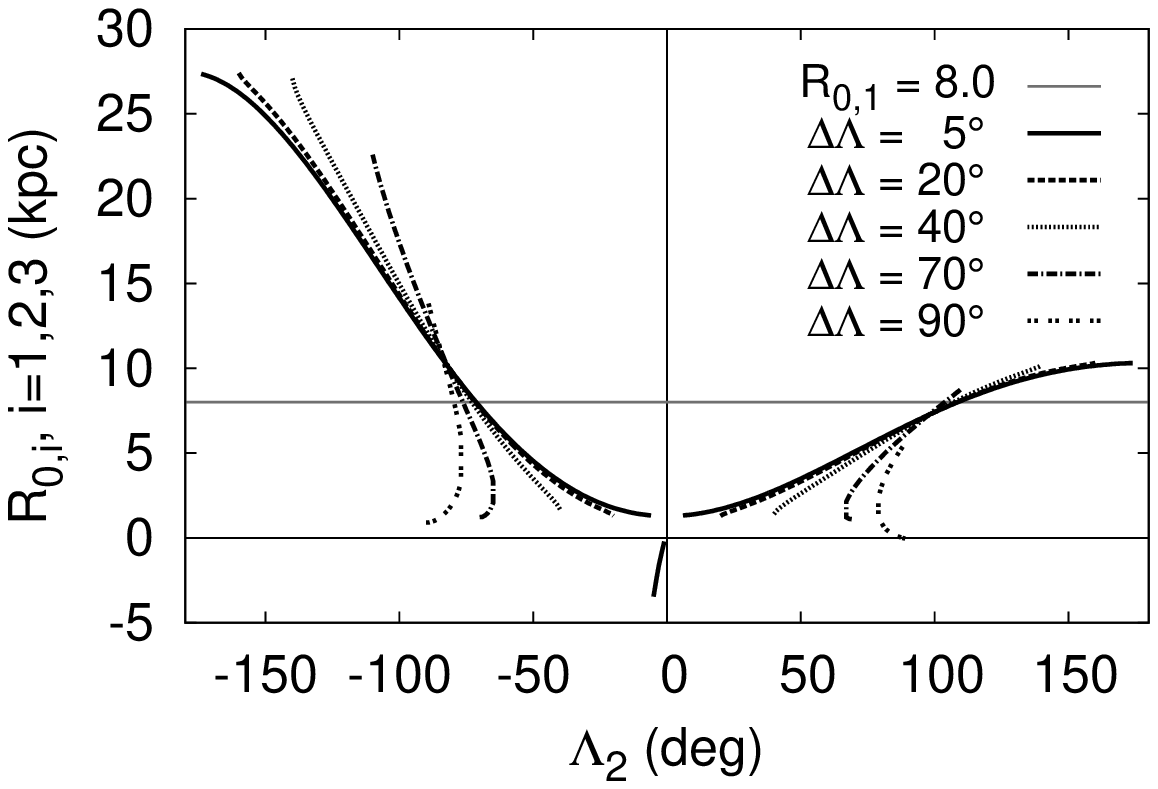,width=88mm,angle=0,clip=}}
\vspace{0mm} \captionb{2}{Dependence of the roots of Equation~(5) on
the longitude of the point $M_2$. Here $\Delta\Lambda$ is the
galactocentric angle between the adjacent points in a triplet. The
gray horizontal line shows the root that is equal to the initial
$R_0$ value, $R_{0,1}=8$~kpc. The curved lines show the branches of
(one or two) additional roots for each  $\Delta\Lambda$ (see text).}
}
\end{figure}

The parameters of the model spiral could be  uniquely determined
from four points belonging to the spiral. In practice, one cannot
draw a one-turn spiral segment through four arbitrary points, and we
therefore apply the three-point method to real and simulated data.
In the cases where the spiral segment is presented by more than
three objects, we estimate the parameters of the segment as the
medians of the sets of values found from all possible triplets of
objects.

\sectionb{3}{APPLICATION OF THREE-POINT METHOD TO MASERS}

We use the catalog of masers by Reid et al. (2014) as a source of
observational data. Ninety masers from this list were assigned by
Reid et al. (2014) to five segments of the Galaxy's spiral arms.

In the cases where most  of the masers of a segment are located on
the same side of the $X$-axis (as for the Scutum and Sagittarius
arms), most of the triplets determine pairs of spirals passing
through the given triplet. In such a situation, we choose the spiral
with the smallest deviation from the segment masers. In the cases
where most of the triplets have a unique spiral passing through all
their three points we exclude the triplets with two possible spirals
from consideration.

We tested various values of the minimum heliocentric angle $\Delta
l_{\mathrm{min}}$ between the adjacent points of the triplet. As
illustrated in Fig.~3, the distribution of the solar galactocentric
distance estimates $R_0$ calculated from triplets of the Perseus arm
has the stable dominant peak and the variance which is dependent on
$\Delta l_{\mathrm{min}}$. We choose the value of $\Delta
l_{\mathrm{min}}$ characterized by the lowest standard error of the
mean of $R_0$ values.


\begin{figure}[!tH]
\vbox{
\hspace*{3pt}
\centerline{\psfig{figure=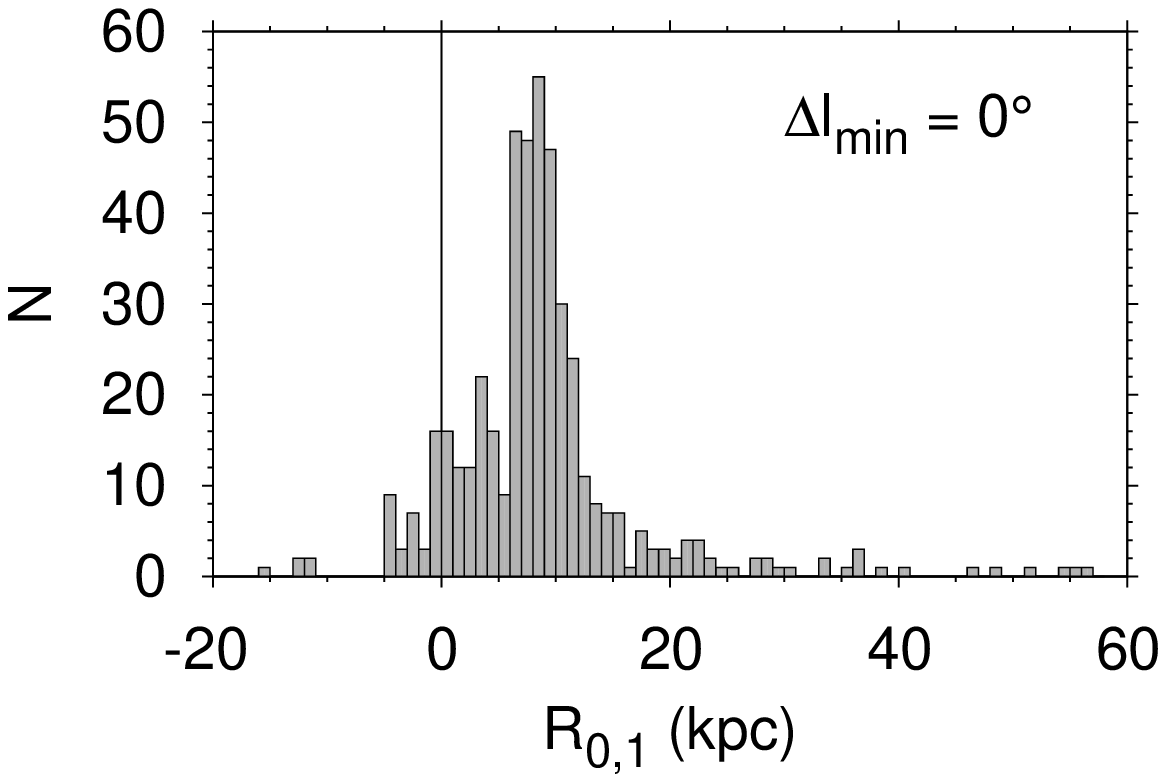,width=61mm,angle=0,clip=}%
\raisebox{0.0pt}{
\hspace*{-2pt}
\psfig{figure=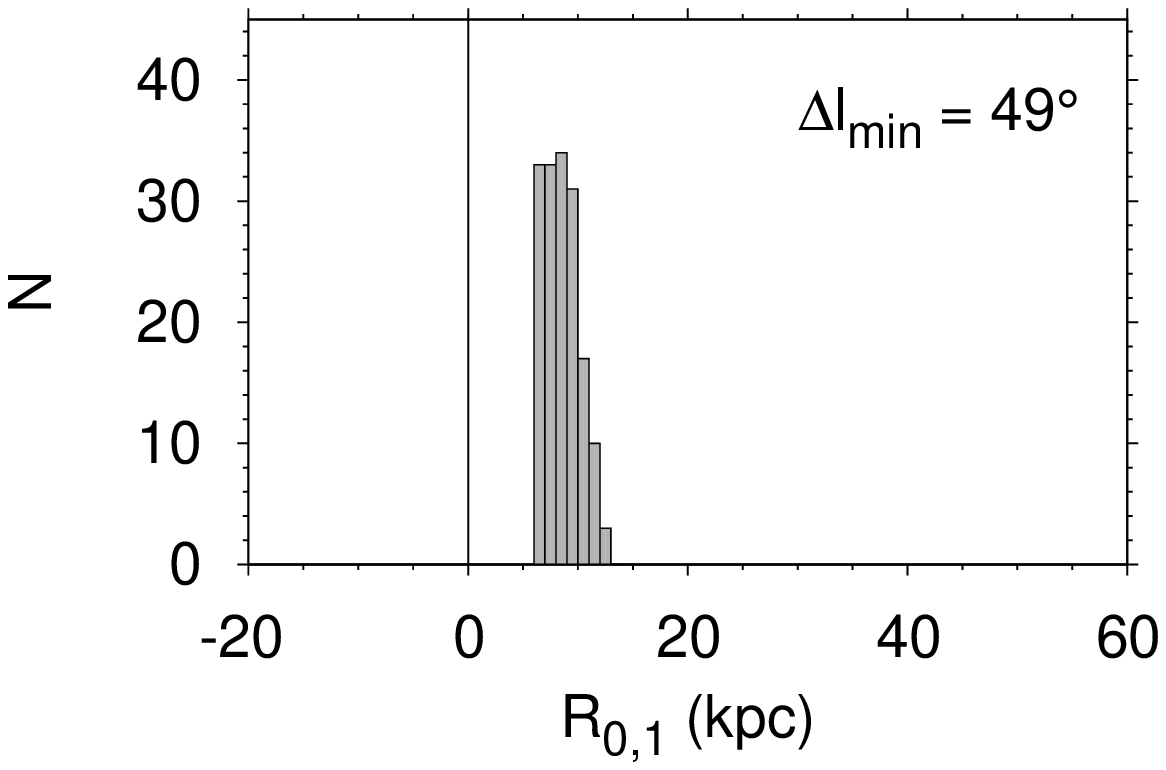,width=61mm,angle=0,clip=}}
}
\captionb{3}{Distribution of the $R_0$ estimates found from all
possible triplets of objects belonging to the Perseus arm.} }
\end{figure}

Table~1 lists the median values, $\me R_0$, of $R_0$ estimates for
five  spiral segments. We used the jackknife technique to compute
the~standard deviations and bias estimates. The weighted average of
the~bias corrected estimates, $R_{0,\mathrm{corr}}$, for the Perseus
and Scutum arms with the weights inversely proportional to the
jackknife variances, $\sigma^2_{R_{0,\mathrm{J}}}$, is $R_0 = 8.44
\pm 0.45$~{kpc}. Three other segments are excluded from
consideration for the following reasons: the Sagittarius arm has
obviously a bimodal distribution of $R_0$ values, the Outer arm has
a very small sample of triplets,  the model spiral for the Local arm
is in rather poor agreement with the position of masers. Hence we
estimate the distance to the Galactic center from masers in spiral
arms to be
\begin{equation}
 R_0 = 8.44 \pm 0.45~\mathrm{kpc}.
\end{equation}
Our simplified three-point method gives an $R_0$ estimate that
agrees with those derived from maser kinematics: $R_0 = 8.34 \pm
0.16~$kpc (Reid et al. 2014) and $R_0 = 8.03 \pm 0.12$~kpc (Bajkova
\& Bobylev 2015).

We also estimated the parameters $i$ and $\lambda_0$ for each
segment as the medians of the sets of values determined from all
possible pairs of objects using Equations~(6) and (7). The estimates
obtained are listed in Table~2 and the resulting spiral structure is
shown in Fig.~4.


\begin{figure}[!]
\vbox{ \centerline{\psfig{figure=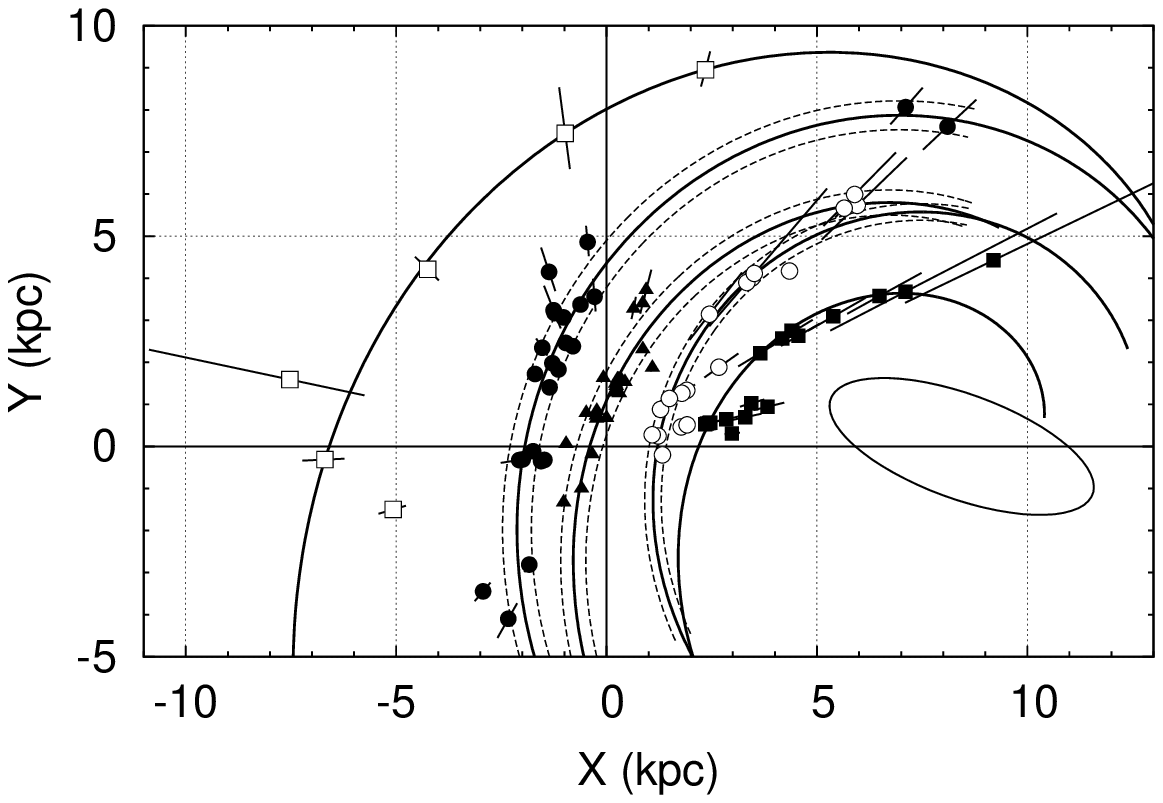,width=90mm,angle=0,clip=}}
\vspace{1pt} \captionb{4}{Spiral pattern and the projected
distribution of masers onto the Galactic plane: Outer arm (open
squares), Perseus arm (solid circles), Local arm (solid triangles),
Sagittarius arm (open circles), and Scutum arm (solid squares). The
galactic bar is shown as an ellipse (J\'\i lkov\'a et al., 2012).} }
\end{figure}


\begin{table}[!]
\begin{center}
\vspace{3mm}\vbox{\footnotesize\tabcolsep=7pt
\parbox[c]{124mm}{\baselineskip=10pt
{\smallbf\ \ Table 1.}{\small\ ~Estimates of $R_0$ for five spiral
segments.  The number of triplets with the~minimum heliocentric
angle $(\Delta l)_{\mathrm{min}}$  is designated as
$N_\therefore$\,.  $\me R_0$ denotes the~median of $R_0$ values
found from all triplets. The $R_{0,\mathrm{corr}}$ and
$\sigma_{R_{0,\mathrm{J}}}$ columns give the~bias corrected solar
galactocentric distance estimate and its standard deviation,
respectively, found using the jackknife technique, and $\Delta
{R_{0,\mathrm{corr}}}$ is the difference between
$R_{0,\mathrm{corr}}$ and $\me R_0$. \hstrut}}
\begin{tabular}{|l|c|c|c|c|c|}
  \hline
   Arm & $N_\therefore$ &  $(\Delta l)_{\mathrm{min}}$ &  $\me R_0^{\vphantom T}$ (kpc)& $ R_{0,\mathrm{corr}} \pm \sigma_{R_{0,\mathrm{J}}} $(kpc) & $\Delta {R_{0,\mathrm{corr}}}$ (kpc)\hstrut\lstrut \\
   \hline
    Outer & 7 & $0^\circ$ &  $8.4^{+5.3^{\vphantom T}}_{-19.6}$ & $16.1 \pm 7.5$ & $+7.7$\\[0.15cm]
   {Perseus} & 161 &$49^\circ$ & $8.43^{+0.19}_{-0.20}$ & ${8.36 \pm 0.53}$ & $-0.07$ \\[0.15cm]
    Local & $328$ & $0^\circ$ & $2.93^{+0.35}_{-0.19}$ & $2.17\pm1.00$ & $-0.76$\\ [0.15cm]
   Sagittarius & 306 & $0^\circ$ & $9.92^{+0.36}_{-0.34}$ & $10.62 \pm 0.69$ &  $+0.70$ \\[0.15cm]
   {Scutum} & 267 & $0^\circ$ & $9.01^{+0.30}_{-0.15}$ & ${8.62 \pm 0.81}$ & $-0.39$\\[0.1cm]
    \hline
  \end{tabular}
}
\end{center}
\end{table}


\begin{table}[!]
\begin{center}
\vbox{\footnotesize\tabcolsep=10pt
\parbox[c]{124mm}{\baselineskip=10pt
{\smallbf\ \ Table 2.}{\small\ ~Estimates of $i$ and $\lambda_0$ for
five spiral segments with $R_0=8.44$~kpc.  Here
${\sigma}_{i,\mathrm{J}}$ and ${\sigma}_{\lambda_0,\mathrm{J}}$ are
the standard deviations calculated using the  jackknife technique,
and $(\sigma_{\mathrm{w}})_{0}$ is the estimated segment's width.
\lstrut}}
 \begin{tabular}{|l|cc|cc|c|}
 \hline
 Arm  & $\me i$ & $ {\sigma}_{i,\mathrm{J}}$ & $\me \lambda_0$ & ${\sigma}_{\lambda_0,\mathrm{J}}$ & $(\sigma_{\mathrm{w}})_{0}$ (kpc) \hstrut\lstrut\\
 \hline
 Outer$^{\vphantom{T^{T^T}}}$   & ${-18\fdg6}^{+6\fdg7}_{-5\fdg6}$ & ${ 0\fdg80}$ & $+98\fdg3^{+25\fdg5}_{-10\fdg8}$ & $ 2\fdg0$  &  \\ [0.15cm]
 Perseus   & ${-10\fdg6}^{+0\fdg55}_{-0\fdg35}$  & ${1\fdg08}$ & $+63\fdg3^{+4\fdg3}_{-2\fdg1}$ & $ 9\fdg4$  & ${0.34} \pm {0.05}$  \\ [0.15cm]
  Local & ${-16\fdg5}^{+1\fdg4}_{-2\fdg2}$ & $ {5\fdg1} $ & $+9\fdg0^{+0\fdg29}_{-0\fdg16}$ & $ 0\fdg55$  & ${0.29} \pm {0.04}$ \\[0.15cm]
 Sagittarius  & ${-9\fdg9}^{+1\fdg8}_{-0\fdg80}$ & $ {3\fdg6}$ & $-50\fdg8^{+6\fdg5}_{-16\fdg7}$ & $26^\circ$ & $ {0.20} \pm {0.04}$ \\ [0.15cm]
 Scutum   & ${-21\fdg4}^{+0\fdg58}_{-1\fdg04}$  & ${1\fdg8}$ & $-43\fdg9^{+2\fdg8}_{-5\fdg7}$ & $ 10\fdg5$ & \\ [0.1cm]
  \hline
 \end{tabular}
}
\end{center}
\end{table}

\sectionb{4}{NUMERICAL INVESTIGATION OF THE STATISTICAL PROPERTIES\\
OF THE $R_0$ ESTIMATES BASED ON SPIRAL SEGMENTS}

\vskip1mm

We use Monte Carlo technique to investigate the statistical
properties of the $R_0$ estimates based on the spiral segment's
geometry. Here we consider a basic model representing the Perseus
arm with the following adopted geometric parameters: $R_0 =
8.0$~kpc, $i = -10\fdg0$, and $\lambda_0 = +61\fdg0$. The
galactocentric longitudes of the segment ends are equal to
$\lambda_1^{\rm s} = -21\fdg0$ and   $\lambda_2^{\rm s} = +88\fdg0$.
We denote the angular length  of  the segment as $\Delta\lambda$.
The simulated segment is populated by $N$=24 objects outlining it
and its width $\sigma_{\mathrm{w}}$  is set equal to  0.34~kpc. The
absolute and fractional parallax errors are set equal to
$\sigma_{\varpi} = 0.025~$mas and  $\sigma_{\varpi}/\varpi = 0.06$,
respectively.  The absolute parallax error varies from maser to
maser  and decreases  with increasing heliocentric distance (see
Fig.~5) and that is why we consider both  absolute and fractional
parallax errors.

We vary one parameter with all other parameters fixed, and
investigate the variation of the~standard deviation and bias of the
estimator. We generated a total of 10\,000 simulated catalogs of
objects for every set of parameters by varying two consequent
offsets: the offset crosswise the arm and the  offset of parallax.
Both offsets are normally distributed with the standard deviations
equal to $\sigma_{\mathrm{w}}$ and $\sigma_{\varpi}$ (or
$\sigma_{\varpi}/\varpi$), respectively.


\begin{figure}[!tH]
\vbox{ \centerline{\psfig{figure=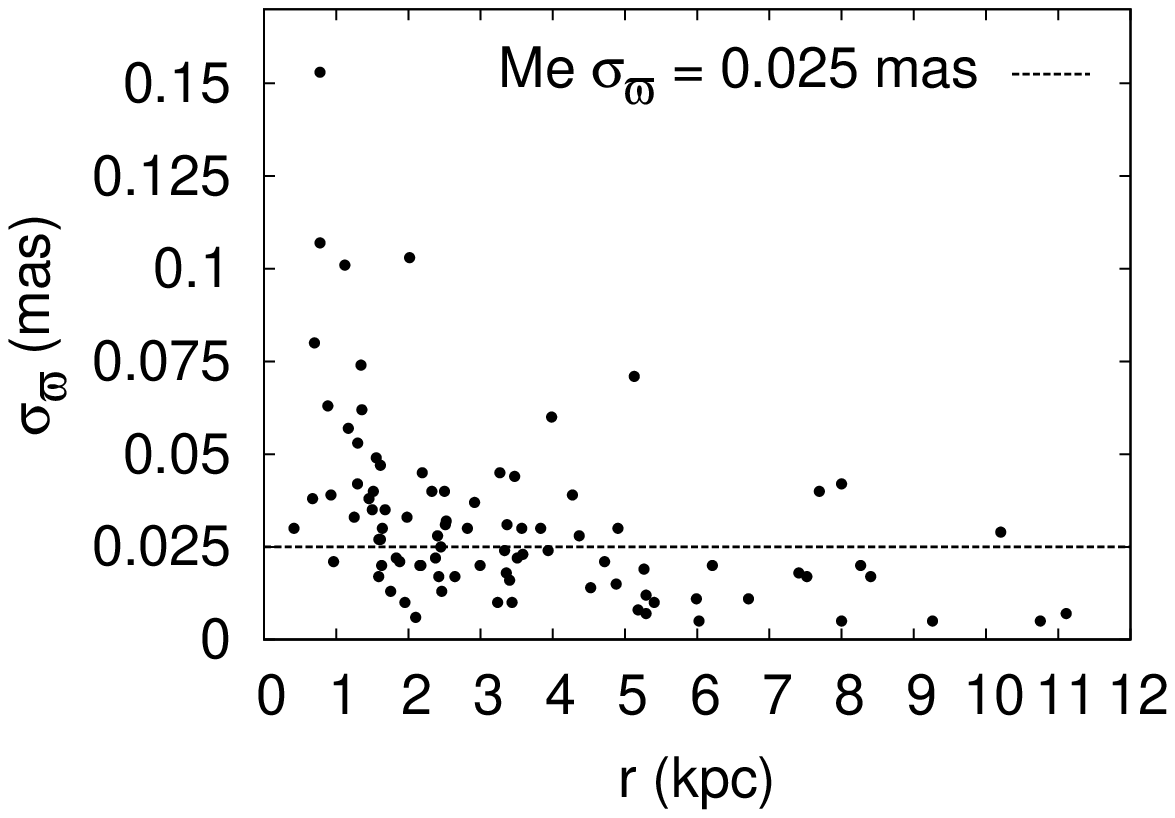,width=80mm,angle=0,clip=}}
\vspace{1pt} \captionb{5}{Dependence of absolute parallax error on
the heliocentric distance to the maser.} }
\end{figure}

Fig.~6 illustrates the dependence of  the median $R_0$, the standard
deviation of $R_0$ estimate, and the uncertainty of $\me R_0$   on
the angular length $\Delta\lambda$ of the segment.  The standard
deviation of $R_0$ decreases by a factor of ten over the
$\Delta\lambda$ interval from $50^\circ$ to $190^\circ$. The bias is
insignificant for the segment lengths $\Delta\lambda$ greater than
$60^\circ$.


\begin{figure}[!tH]
\vbox{
\hspace*{5pt}
\centerline{\psfig{figure=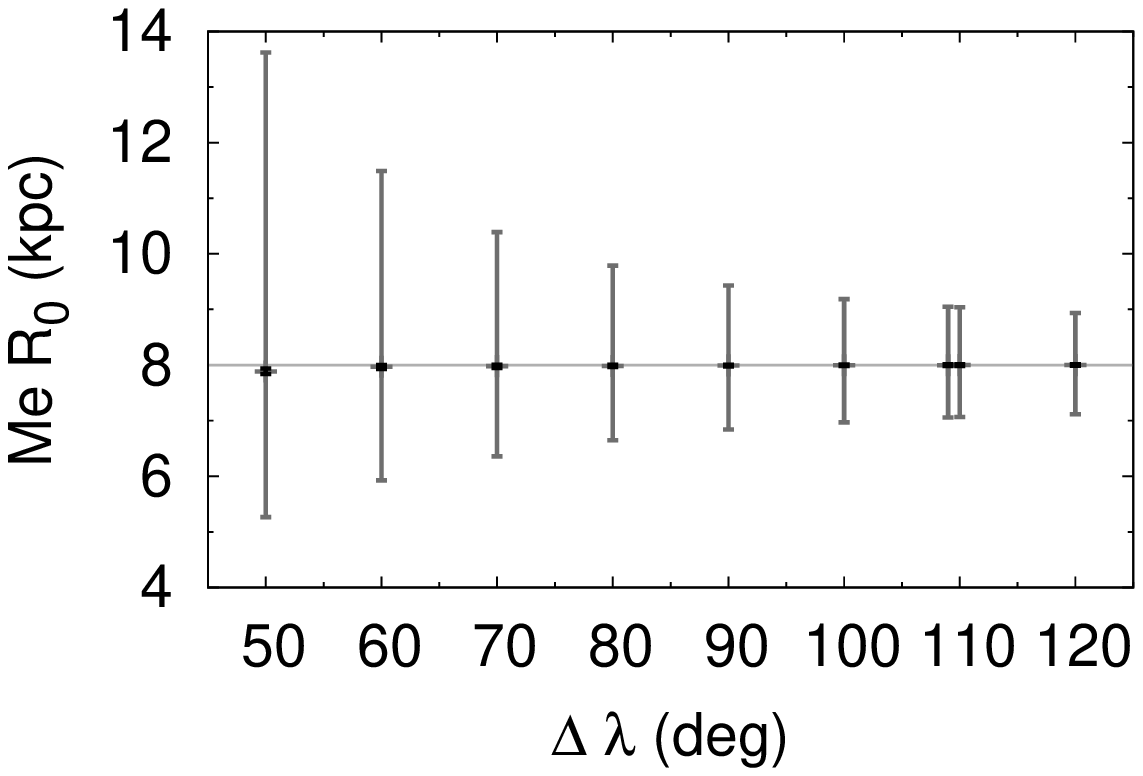,width=60mm,angle=0,clip=}%
\raisebox{0.0pt}{
\hspace*{0pt}
\psfig{figure=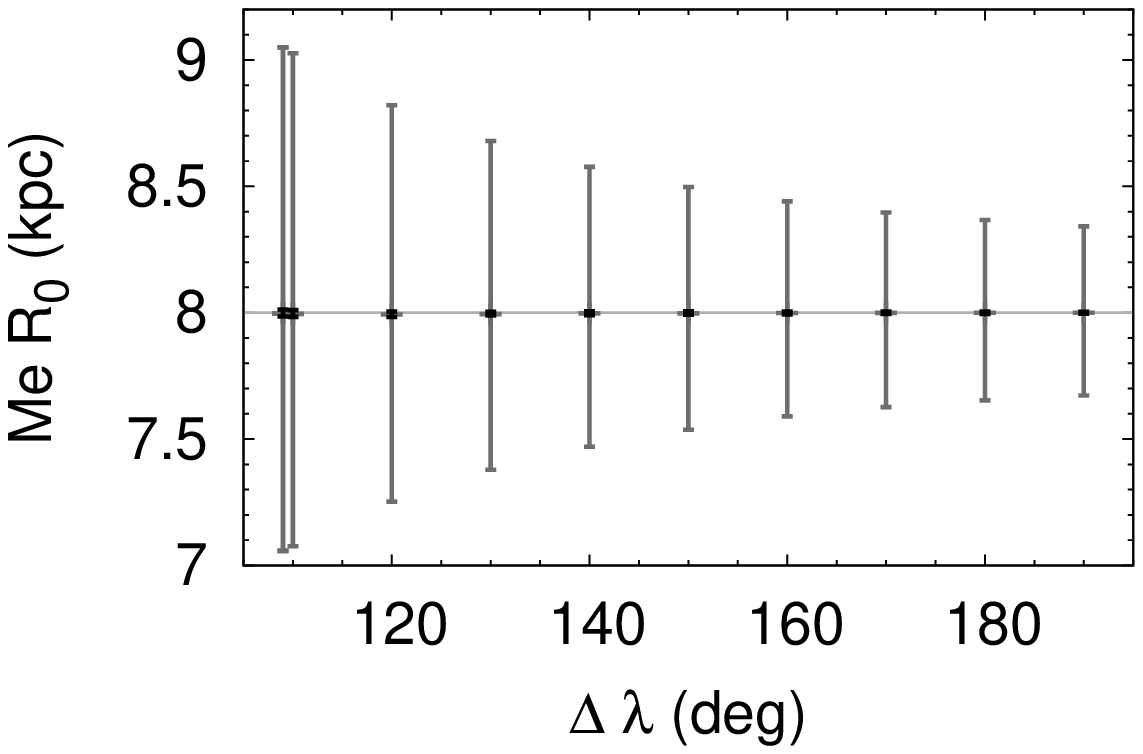,width=60mm,angle=0,clip=}}
}
\captionb{6}{Median and standard deviation of $R_0$ estimate (the
gray dashes and error bars) and the uncertainty of $\me R_0$ (the
black error bars) vs.  $\Delta\lambda$ in the case of constant
$\lambda_1^{\rm s}$ (left panel) and $\lambda_2^{\rm s}$ (right
panel).} }
\end{figure}

Fig.~7 shows the dependence of the same statistical characteristics
on the absolute ($\sigma_{\varpi}$) and fractional
($\sigma_{\varpi}/\varpi$) parallax error. The standard deviation of
the $R_0$ estimate increases by a factor of three over the interval
from $\sigma_{\varpi}$~=~$0$ to $0.05$~mas.


\begin{figure}[!tH]
\vbox{
\hspace*{5pt}
\centerline{\psfig{figure=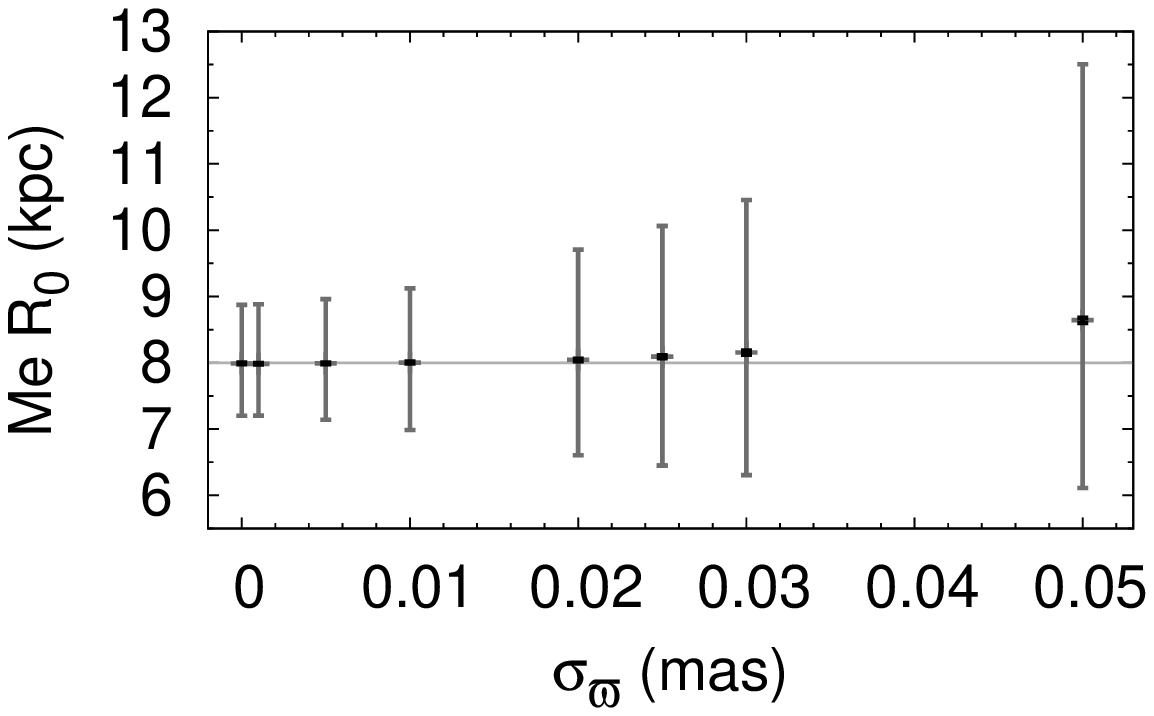,width=60mm,angle=0,clip=}%
\raisebox{0.0pt}{
\hspace*{0pt}
\psfig{figure=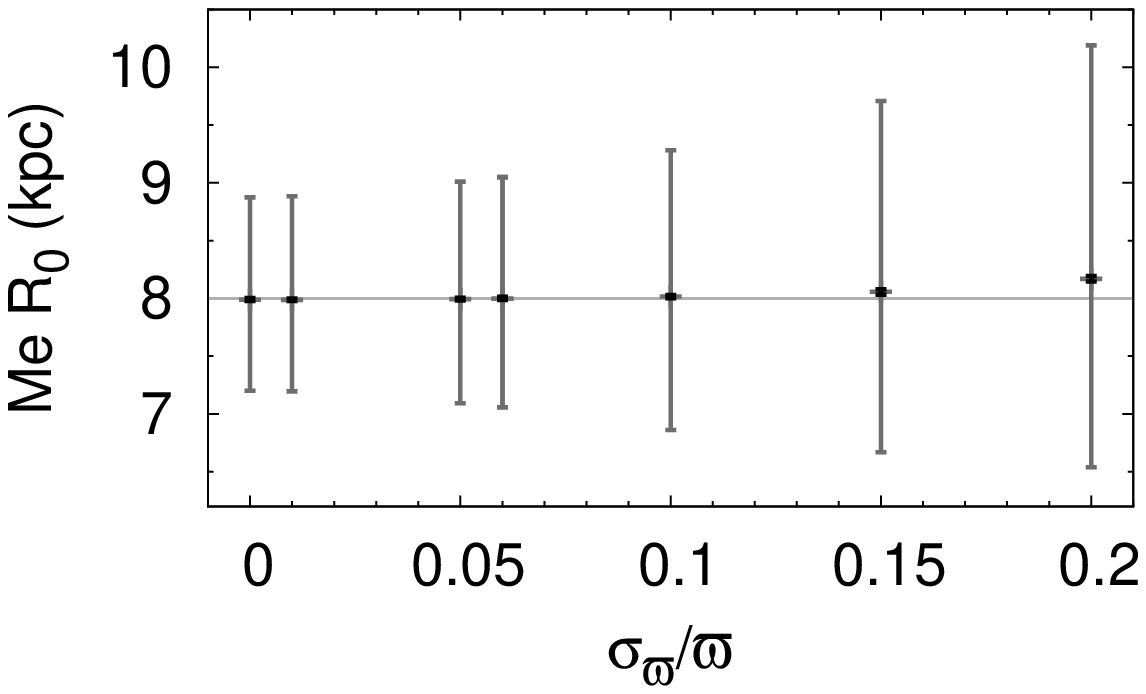,width=60mm,angle=0,clip=}}
}
\captionb{7}{Dependence of the median and standard deviation of
$R_0$ estimate (the gray dashes and error bars) and the uncertainty
of $\me R_0$ (the black error bars)  on absolute
parallax error $\sigma_{\varpi}$ (left panel) and fractional parallax error $\sigma_{\varpi}/\varpi$ (right panel).}
}
\end{figure}

The largest biases are found in the cases of  large uncertainty
$\sigma_{\varpi}$ and small number of objects in the segment (see
Table~3). The standard deviation is mostly influenced by
the angular length $\Delta\lambda$ and absolute uncertainty of parallax
$\sigma_{\varpi}$.


\begin{table}[!t]
\begin{center}
\vspace{2mm} \vbox{\footnotesize\tabcolsep=8pt
\parbox[c]{124mm}{\baselineskip=10pt
{\smallbf\ \ Table 3.}{\small\  Results of the numerical
simulations. Each parameter $p$ varies from $p_{\mathrm{min}}$
to~$p_{\mathrm{max}}$. Here $\sigma_{R_0}$ and $\Delta R_0$ denote
the standard deviation of the $R_0$ estimate and the bias of the
estimator, respectively.\lstrut}}
 \begin{tabular}{|c||c|ll||c|ll|}
 \hline
$p$ & $p_{\mathrm{min}}$ &  $\sigma_{R_0}$       & $\Delta R_0$       &  $p_{\mathrm{max}}$ &  $\sigma_{ R_0}$        & $\Delta R_0$      \\
    &                    &                 (kpc) &              (kpc) &                     &                  (kpc)  &              (kpc)\\
   \hline
   $\Delta \lambda$, $\lambda_1^s = -21^\circ$ & $50^\circ$  & $ {}_{-2.6}^{+5.7}  $ & $-0.12^{\vphantom{T^{T^T}}} \pm 0.05$ &  $120^\circ$ & ${}_{-0.89}^{+0.94}$ & $-0.00 \pm 0.01$\\ [0.1cm]
   $\Delta \lambda$,  $\lambda_2^s = +88^\circ$  & $109^\circ$  & $ {}_{-0.94}^{+1.05}  $ & $-0.00_{-0.01}^{+0.02}$ &  $190^\circ$ & ${}_{-0.33}^{+0.34}$ & $-0.00 \pm 0.04$\\ [0.1cm]
  $\sigma_{\varpi}$ (mas) & $0.00$ & $ {}_{-0.79}^{+0.89}$ & $+0.01\pm 0.01$ & $0.05$ & $ {}_{-2.5}^{+3.9}$ & $ {+0.65}_{-0.05}^{+0.03}$ \\ [0.1cm]
       $N$    & $3$ & $ {}_{-2.1}^{+3.3}$ &  $  {+0.61}\pm 0.03$ & $60$ &  $ {}_{-0.61}^{+0.66}$ &  $+0.00\pm 0.01$\\ [0.1cm]
     $\sigma_{\mathrm{w}}$ (kpc) & $0.00$ & $ {}_{-0.52}^{+0.54}$ & $+0.00 \pm 0.01$ & $0.60$ &  $ {}_{-1.3}^{+1.7}$ & $+0.04 \pm 0.02$  \\[0.1cm]
   $\sigma_{\varpi}/\varpi$ & $0.00$ & $ {}_{-0.79}^{+0.89}$ & $+0.01\pm 0.01$  & $0.20$ & $ {}_{-1.6}^{+2.0}$ & $+0.17 \pm 0.02$  \\[0.1cm]
      $i$ & $-20^\circ$  & $ {}_{-0.85}^{+0.94}$& $+0.00 \pm 0.01$  & $0^\circ$ & $ {}_{-0.96}^{+1.10}$  & $-0.02_{-0.01}^{+0.02}$\\ [0.1cm]
    \hline
 \end{tabular}
}
\end{center}
\end{table}

\sectionb{5}{CONCLUSIONS}

We have for the first time determined the solar galactocentric
distance $R_0$ from the spatial distribution of objects tracing the
spiral arms of the Galaxy. We used the data for masers with measured
trigonometric parallaxes to yield an estimate of $R_0 = 8.44 \pm
0.45$~{kpc}. Parameters of five Galactic spiral segments were
evaluated. The results of numerical simulations support the
efficiency of our new approach.

\thanks{The authors acknowledge the support from the Saint Petersburg
State University (research grant No.~6.37.341.2015).}

\References

\refb Bajkova A. T., Bobylev V. V. 2015, Baltic Astronomy, 24, 43

\refb Bobylev V. V., Bajkova A. T.  2014, MNRAS, 437, 1549

\refb Efremov Yu. N. 2011, Astronomy Reports, 55, 108

\refb Francis C., Anderson E. 2012,  MNRAS, 422, 1283

\refb  J\'\i lkov\'a L., Carraro G., Jungwiert B., Minchev I. 2012, A\&A, 541, AA64

\refb Nikiforov, I.I. 2004, in {\it Order and Chaos in Stellar and
Planetary Systems}, eds. G. G. Byrd et al.,  ASP Conf. Ser., 316,
199

\refb Nikiforov I. I., Shekhovtsova T. V. 2001, in {\it Stellar
Dynamics: from Classic  to Modern}, eds. L. P. Ossipkov \& I. I.
Nikiforov, St.~Petersburg Univ. Press, 28

\refb Reid M. J., Menten K. M., Brunthaler A. et al. 2014, ApJ, 783, 130

\refb Xu Y.,  Li J. J., Reid M. J. et al. 2013, ApJ, 769, 15

\end{document}